\documentclass[aps,prl,superscriptaddress,showpacs,twocolumn]{revtex4}
\usepackage{epsfig}
\begin{document}
\title{Anti-Kondo regime of charge transport through a double dot molecule}
\author{Richard Berkovits}
\affiliation{The Minerva Center, Department of Physics,
    Bar-Ilan University, Ramat-Gan 52900, Israel}
\author{Boris Altshuler}
\affiliation{Physics Department, Columbia University, New York, NY 10027}
\date{September 29, 2006, version 2.1}

\begin{abstract}
The conductance through a serial double dot structure for which the inter-dot tunneling
is stronger than the tunneling to the leads is studied using the numerical density matrix
renormalization group method and analytic arguments. When the dots are occupied by 
$1$ or $3$ electrons the usual Kondo peak is obtained. For the case in which $2$ electrons 
occupy the molecule a singlet is formed. Nevertheless, the conductance in that case has a 
constant non-zero value, and might even be equal to the maximum conductance
of $2 e^2/h$ for certain values
of the molecule parameters. We show that this is the result of the subtle interplay
between the symmetric and anti-symmetric orbitals of the molecule caused by interactions
and interference.

\end{abstract}
\pacs{73.23.Hk,71.15.Dx,73.23.-b }

\maketitle

Double quantum dot devices have recently drawn much attention both for
their relevance to possible technological applications such as qubits
\cite{craig04,mason04,koppens05,petta05,sampaz06}, 
as well as the light they shed on basic concepts such as
the Kondo effect and the Fano resonance \cite{jeong01,craig04,johnson04,kobayashi04}.
The double quantum dot molecule can be connected in parallel to the leads, such that
electrons may tunnel from each dot to any lead. In this case one expects
interference between paths going through each of the dots to play an important
role in the transport through the molecule. The interplay between interference
and interaction effects such as Coulomb blockade, spin and orbital Kondo effects,
is the subject of many recent papers \cite{craig04,mason04,koppens05,petta05,sampaz06,jeong01,johnson04,kobayashi04,jones88,mravlje06,kuznetsov96,yacoby95}.

In comparison, the transport through a
serially connected molecule, for which electrons from the right
lead can tunnel only to the right dot and vise versa, is expected to
be much duller. At very low temperatures one anticipates that for an odd
filling of a molecule composed of two identical dots 
the Kondo effect will dominate the conductance.
If the tunneling between the dots is much weaker than the coupling of the
dot to the lead an orbital Kondo effect (where the two degenerate states are
an electron occupying the left or right dot) is expected. In the opposite limit
a spin Kondo effect will be observed
(where the degenerate states are a spin up or down electron
occupying the symmetric or anti-symmetric superposition of both dots orbitals).
For a certain value of intermediate tunneling between the dots a SU(4) 
Kondo effect is expected. For even occupation of the molecule one may expect
the usual Coulomb 
blockade scenario, i.e., once the number of electron on the molecule
is integer no conduction through it is expected. 

This naive picture may
be broken by the following scenario: 
Each of the two electrons is localized on a single dot and forms
a Kondo singlet with the electrons of the corresponding leads \cite{jones88}.
Transport is then possible between the left and right Kondo state, and the conductance
may even reach the maximum value of $2 e^2/h$ for certain values 
of the molecule parameters \cite{mravlje06}. This state can be energetically
favorable only if the gain from the formation of a Kondo state (which is of order
of the Kondo temperature $T_K$) is larger than the loss of kinetic and 
interaction energy due to the localization of the electron on a single dot (i.e.,
the triplet-singlet energy separation). Since $T_K$ depends exponentially on the coupling
of the dots to the leads, one expects this behavior only if the hopping between the
dot and the lead is much larger than the inter-dot hopping. 

In this paper we shall show that also in the opposite limit where the inter-dot hopping
is much larger than the hopping to the leads, finite conductance for the doubly
occupied molecule of two serial dots is possible and
may even reach the maximum value of $2 e^2/h$. Since in this state the ground
state of the molecule is a singlet, this conductance is not connected to the Kondo effect.
Rather, this conductance stems from interactions in the molecule, which
result in partial occupation of  both the symmetric and anti-symmetric states. 
Thus both states can carry current even when the average occupation of the
molecule is integer. It should be noted however that this current vanishes
in the limit when the symmetric and anti-symmetric states are degenerated
- this limit corresponds to two disconnected dots. On the other hand in
the opposite limit when the energy separation is large the situation is
similar to the conventional Coulomb Blockade, i.e., the current should be
small for any integer occupation. All in all there should be some optimal
splitting between the two states, which corresponds  to a maximum of the
conductance. As we shall show the non-zero
conductance valley
is constant for a wide region of gate voltages and is only weakly dependent on
the temperature and external magnetic field, which is very different than the
typical Kondo behavior.

The serial double quantum dot molecule model is defined by the Hamiltonian
\begin{equation}
H = H_{\rm molecule} + H_{\rm leads} + H_{\rm mix}.
\label{hamiltonian}
\end{equation}
The Hamiltonian, $H_{\rm molecule}$, of the double dot molecule is given by
\begin{eqnarray}
H_{\rm molecule} =  \sum_{i=1,2;\sigma=\uparrow,\downarrow}
(\epsilon + h \sigma - \mu) b_{i \sigma}^\dagger b_{i \sigma} 
+\sum_{\sigma} t b_{1 \sigma}^\dagger b_{2 \sigma} + {\rm h.c.} \nonumber \\
+ U \sum_{i=1,2} b_{i \uparrow}^\dagger b_{i \uparrow} 
b_{i \downarrow}^\dagger b_{i \downarrow}
+ U^\prime \sum_{\sigma,\sigma^\prime=\uparrow,\downarrow} 
b_{1 \sigma}^\dagger b_{1 \sigma} b_{2 \sigma^\prime}^\dagger b_{2 \sigma^\prime},
\label{hamiltonian_mol}
\end{eqnarray}
where the energy level $\epsilon$ of each dot (with creation operator
$b_{1(2)\sigma}^\dagger$ for an electron in the first (second) dot with spin $\sigma$), 
the intra-dot hopping matrix element $t$ and the charging energies $U$ 
for each dot and the
mutual capacitive coupling energy $U^\prime$. 
By applying a gate voltage $V_g$ to the molecule one may change the dot's level $\epsilon$
to $\epsilon-V_g$.
In principle also the hopping between the dots, $t$, may be changed by applying a
side gate (as well as the hopping between the dot and the lead).
For any realistic configuration
one expects $U>U^\prime$. The influence of
an external magnetic field $h$ and chemical potential $\mu$ are 
included in the first term of Eq. (\ref{hamiltonian_mol}).
The lead Hamiltonian,
$H_{\rm leads} = \sum_{j,\sigma=\uparrow,\downarrow,\lambda=L,R} 
-\mu a_{j \sigma \lambda}^{\dagger} a_{j \sigma \lambda} +
(a_{j \sigma \lambda}^{\dagger} a_{j+1 \sigma \lambda} + {\rm h.c.})$ 
and the tunneling between lead and molecule is described by
$H_{\rm mix} = V \sum_{\sigma=\uparrow,\downarrow} 
(a_{1 \sigma L}^\dagger b_{1 \sigma} +
a_{1 \sigma R}^\dagger b_{2 \sigma})+{\rm h.c.}$. Here $a_{j \sigma \lambda}^{\dagger}$
is the creation operator in the i-th site of the $\lambda=L(R)$ left (right) lead,
the hopping in the leads is set to one and the hopping between the lead and dot
is $V$.

It is useful to first consider the eigenvalues and eigenvectors of a disconnected
molecule, i.e., to diagonalize $H_{\rm molecule}$. 
Defining the single particle basis as
$|+\rangle = (|1\rangle+|2\rangle)/\sqrt{2}$ and 
$|-\rangle = (|1\rangle-|2\rangle)/\sqrt{2}$ 
(where $|1\rangle,|2\rangle$ are the orbitals
of the first and second dot), one can denote the many particle Hilbert space
by the application of the creation operators
$d_{+ \sigma}^\dagger=(b_{1 \sigma}^\dagger - b_{2 \sigma}^\dagger)/\sqrt{2})$
and $d_{- \sigma}^\dagger=(b_{1 \sigma}^\dagger - b_{2 \sigma}^\dagger)/\sqrt{2})$
on the vacuum $|vac\rangle$. Thus, for example,  $|\uparrow,\downarrow\rangle
=d_{+\uparrow}^\dagger d_{-\downarrow}^\dagger |vac\rangle$ and
$|\uparrow\downarrow,0\rangle
=d_{+\uparrow}^\dagger d_{+\downarrow}^\dagger |vac\rangle$.
The matrix corresponding to $H_{\rm molecule}$ is block-diagonal with respect to
the number of electrons $N$ ($N$ may vary between zero and four) and each
block may be diagonalized independently. The following lowest eigenvalues and
eigenstates for each block are found: $\varepsilon(N=0)=0,|vac\rangle$ ;
$\varepsilon(N=1)=\epsilon-\mu-t ; |\uparrow,0\rangle$ and
$\varepsilon(N=1)=\epsilon-\mu-t ; |\downarrow,0\rangle$ (double degeneracy) ;
$\varepsilon(N=2)=2(\epsilon-\mu)+(U+U^\prime)/2-\sqrt{(U-U^\prime)^2
+(4t)^2}/2, 
(1+[\sqrt{1+[4t/(U-U^\prime)]^2}-4t/(U-U^\prime)]^2)^{-1}
\{|\uparrow \downarrow,0\rangle + 
(\sqrt{1+[4t/(U-U^\prime)]^2}-4t/(U-U^\prime))|0,\uparrow \downarrow\rangle\}$ ;
$\varepsilon(N=3)=3(\epsilon-\mu)+U+2U^\prime-t ; |\uparrow \downarrow,\uparrow\rangle$ and
$\varepsilon(N=3)=3(\epsilon-\mu)+U+2U^\prime-t ; |\uparrow \downarrow,\downarrow\rangle$ 
(double degeneracy) ; and $\varepsilon(N=4)=4(\epsilon-\mu)+2U+4U^\prime,
|\uparrow \downarrow,\uparrow \downarrow\rangle$.

Generally, one expects that coupling the molecule to the leads will
cause broadening of the molecule states, but as long as $V<t$ the state
of the molecule will continue to retain their identity. Thus one might
predict the following behavior of the molecules occupation (i.e.,
number of electrons on the molecule) an function of $\epsilon$.
For $\epsilon>\mu-t$ the dot population $N=0$. Around $\epsilon \sim \mu-t$
the population switches to $N=1$. At
$\epsilon \sim \mu-(U+U^\prime)/2+\sqrt{(U-U^\prime)^2
+(4t)^2}/2-t$ the population increases to $N=2$, while at
$\epsilon \sim \mu-U/2-3U^\prime/2-\sqrt{(U-U^\prime)^2
+(4t)^2}/2+t$ it switches to $N=3$. Finally at 
$\epsilon \sim \mu-U-2U^\prime-t$ the molecule will be fully occupied.

According to the ''orthodox'' theory \cite{alhassid00} one would
expect a peak in the conductance each time when the population switches.
Thus one would expect four peaks in the conductance as a function of $\epsilon$ at
the values of the switches in the molecule population given previously.
This consideration will not hold at zero temperature and zero magnetic
field ($T,h=0$), provided that Kondo physics will play an important role.
Thus for odd occupancy of the molecule ($N=1,3$), for which the ground state
is degenerate, we expect to see a Kondo plateau in which the conductance is equal
to $2 e^2/h$. On the other hand for the $N=2$ occupancy 
there are no degeneracies in the molecule ground state, e.g., the singlet 
($S=0$) ground state
is quite far away from the triplet state.
Therefore naively
one expects no conductance. Thus, at $T=0$ we expect two broad
conduction peaks separated by a zero conductance valley.
The width of the Kondo conductance peaks are $2t+(U+U^\prime)/2-\sqrt{(U-U^\prime)^2
+(4t)^2}/2$, while the width
of the valley is $-2t+U^\prime+\sqrt{(U-U^\prime)^2
+(4t)^2}$. 

\begin{figure}\centering
\epsfxsize7cm\epsfbox{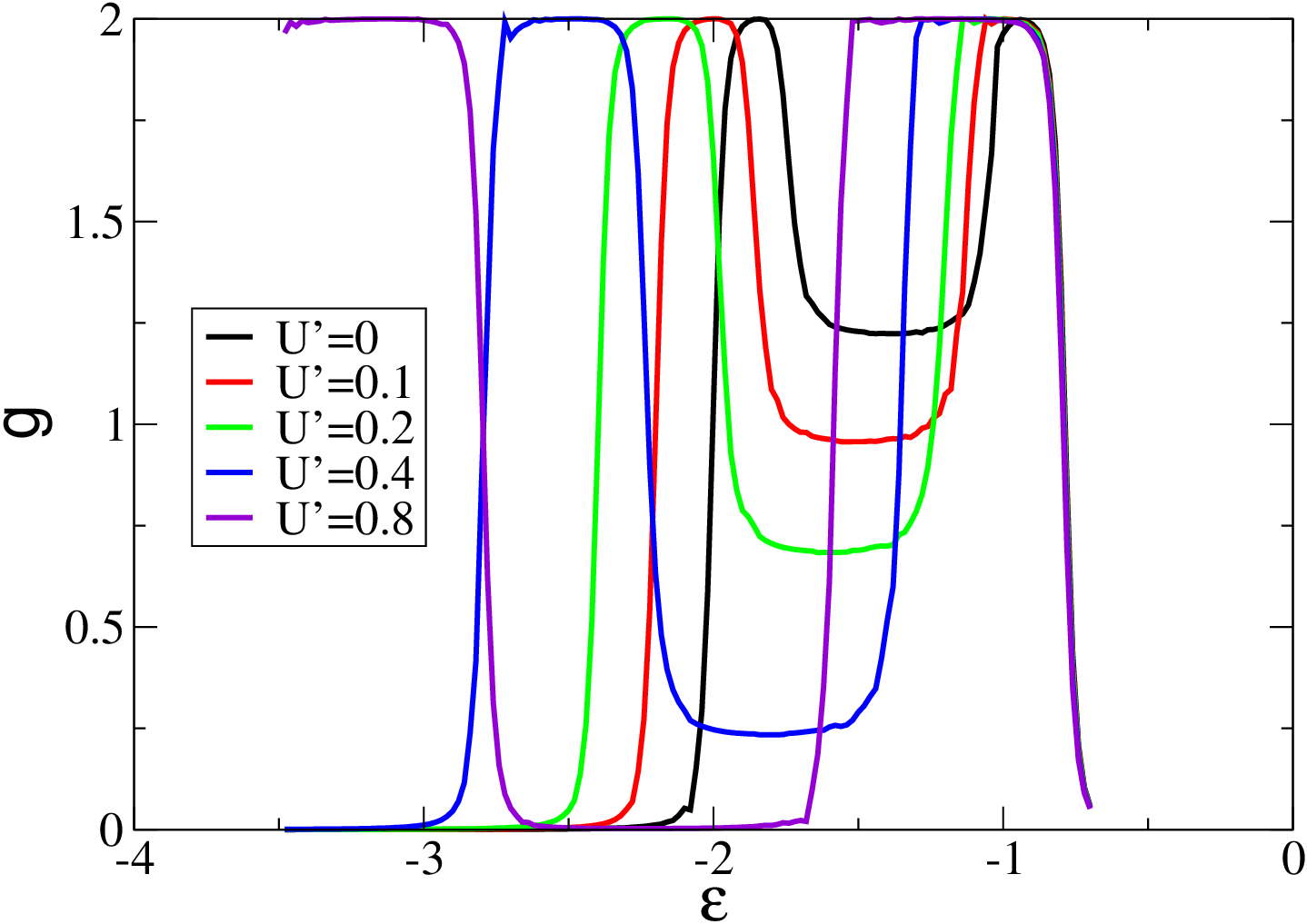} \vskip .8truecm 
\epsfxsize7cm\epsfbox{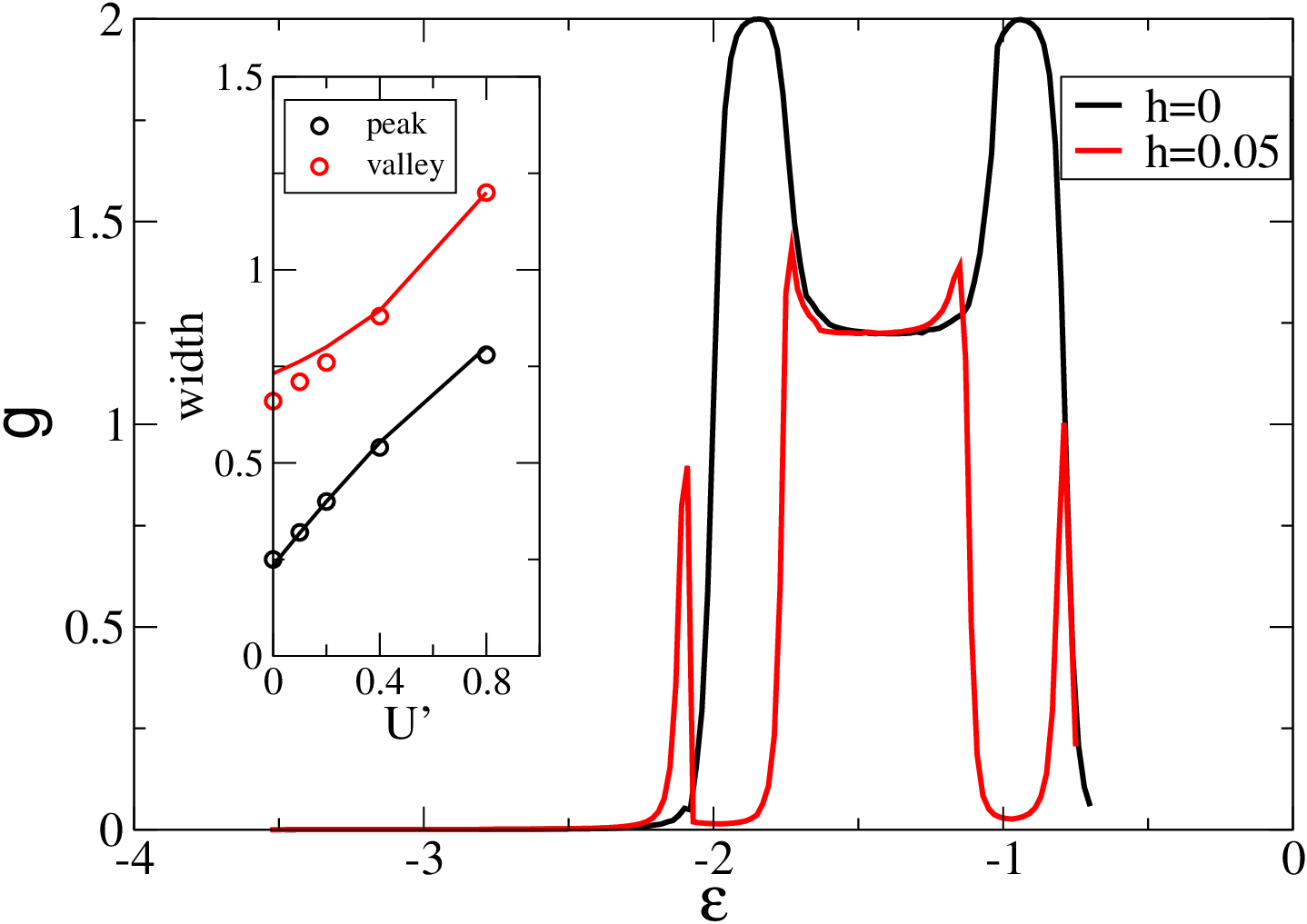} \vskip -.3truecm 
\caption{
(a) Conductance, $g$, vs. gate voltage (level position), $\epsilon$,
for different values of the inter-dot interaction $U^\prime$, while
the on-site interaction $U=0.8$, inter-dot hopping $t=0.2$ and dot-lead coupling
$V=0.1$ are kept constant and no external magnetic field is applied..
(b) $g$, vs. $\epsilon$ in the presence of a magnetic field $h=0.05$
compared to $g$ in the absence of such a field. Here $U^\prime=0$,
and the other parameters are as in (a). Inset: the peak and valley width
as function of $U^\prime$. Symbols correspond to the numerical results
while lines to theory.
} \label{fig1}
\end{figure}

Once we attach the molecule to the leads, the
problem becomes much more complicated.We compute the
ground state for the interacting molecule attached to a couple of 1D leads 
using an extension of a DMRG method previously used to calculate the ground 
state of a dot attached to leads \cite{berkovits04}. 
The essence of the method is similar to the regular DMRG for
1D systems \cite{white93}. Once we obtain the many-body ground state eigenvector  
$|0\rangle$ of the
entire system we can calculate the occupation of the $|+\rangle$ and  $|-\rangle$
orbitals of the molecule defined as $n_{\pm \sigma} = 
\langle 0| d_{\pm\sigma}^\dagger d_{\pm\sigma} |0 \rangle$. 
For the case of the double quantum 
dot with each of the two dots' is coupled symmetrically to a corresponding lead,
and the interactions don't explicitly break the symmetry, 
the Friedel sum rule implies 
the relation $g=\sin^2(\pi(n_{+\uparrow}-n_{-\uparrow}))+
\sin^2(\pi(n_{+\downarrow}-n_{-\downarrow}))$ in the usual way \cite{datta97}. 
Thus, the occupations $n_{\pm \sigma}$ determines the conductance.

In Fig. \ref{fig1}a we present the numerical results obtained for the conductance as
function of the gate voltage ($\epsilon$) for different values of
the inter-dot interaction $U^\prime$, while $U=0.8$,$t=0.2$ and $V=0.1$ are kept constant.
The two broad Kondo peaks (for which $g=2$) are very clear and their width
corresponds to our expectation $2t+(U+U^\prime)/2-\sqrt{(U-U^\prime)^2
+(4t)^2}/2$ (see the inset of Fig. \ref{fig1}b) obtained for a disconnected molecule.
The coupling with the leads rounds the peaks, but does not significantly change
their width. Thus, in the absence of
interactions ($U,U^\prime=0$) there is no peak (the width is equal to zero), while
for $U=U^\prime$ the width is simply $U$.

The surprise in Fig. \ref{fig1}a comes from the behavior of the valley. Although its
width corresponds quite well to the expectation of $-2t+U^\prime+\sqrt{(U-U^\prime)^2
+(4t)^2}$ (see the inset of Fig. \ref{fig1}b) the conductance in the valley is constant
and not necessarily equal to zero. The $N=2$ state conductance is not suppressed by
applying a magnetic field. As seen in Fig. \ref{fig1}b, applying a magnetic
field that is enough to suppress the Kondo peaks of the $N=1$ and $N=3$ sectors,
resulting in the expected four conductance peaks, does
not change much the conductance at the $N=2$ valley. This is expected since
the ground state of the disconnected molecule for the $N=2$ sector is a singlet and
does not couple to the external magnetic field. Only for a magnetic field
$h>(U^\prime-U)/2+\sqrt{(U-U^\prime)^2+(4t)^2}/2$ will the triplet excited state
($|\uparrow\uparrow,0\rangle$) cross the singlet state and the conductance
will drop to zero. Although the conductance in the valley has some superficial
similarities to the Kondo conductance (i.e., is constant for a wide range of gate
voltages), the fact that the ground state in the molecule has no degeneracies rules
out any kind of Kondo-like scenarios. 

In fact a wide region of almost constant conductance for double dots connected in
series could be seen in the data presented in studies using slave boson formalism
\cite{aono98}, and numerical renormalization methods \cite{izumida00}. Hints of
this behavior are also present in the study for 3 dots in the $N=2$ and $N=4$ 
regions \cite{oguri05}. Nevertheless, the straight-forward expectation is that the 
conductance of the $N=2$ region of the double dot molecule is zero unless some 
ferromagnetic coupling between the dots will create a triplet ground state  
for $N=2$ \cite{georges99,martins05}, or when each dot is more strongly coupled to its
lead than to each other, in which case a separate left and right Kondo state
form and transport between these states is possible \cite{jones88}. Here none of
these explanations is relevant.

\begin{figure}\centering
\epsfxsize7cm\epsfbox{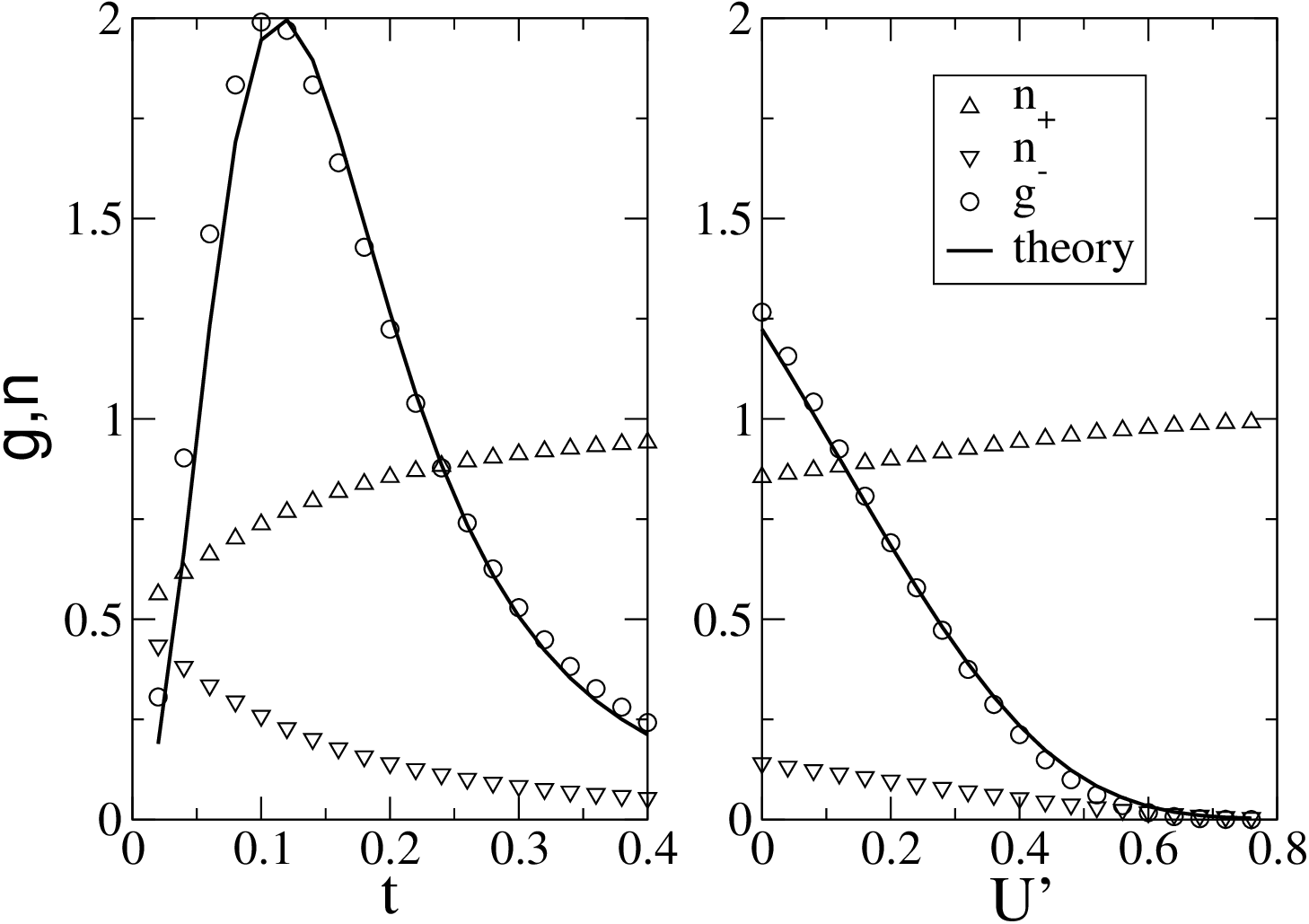} \vskip .8truecm 
\epsfxsize7cm\epsfbox{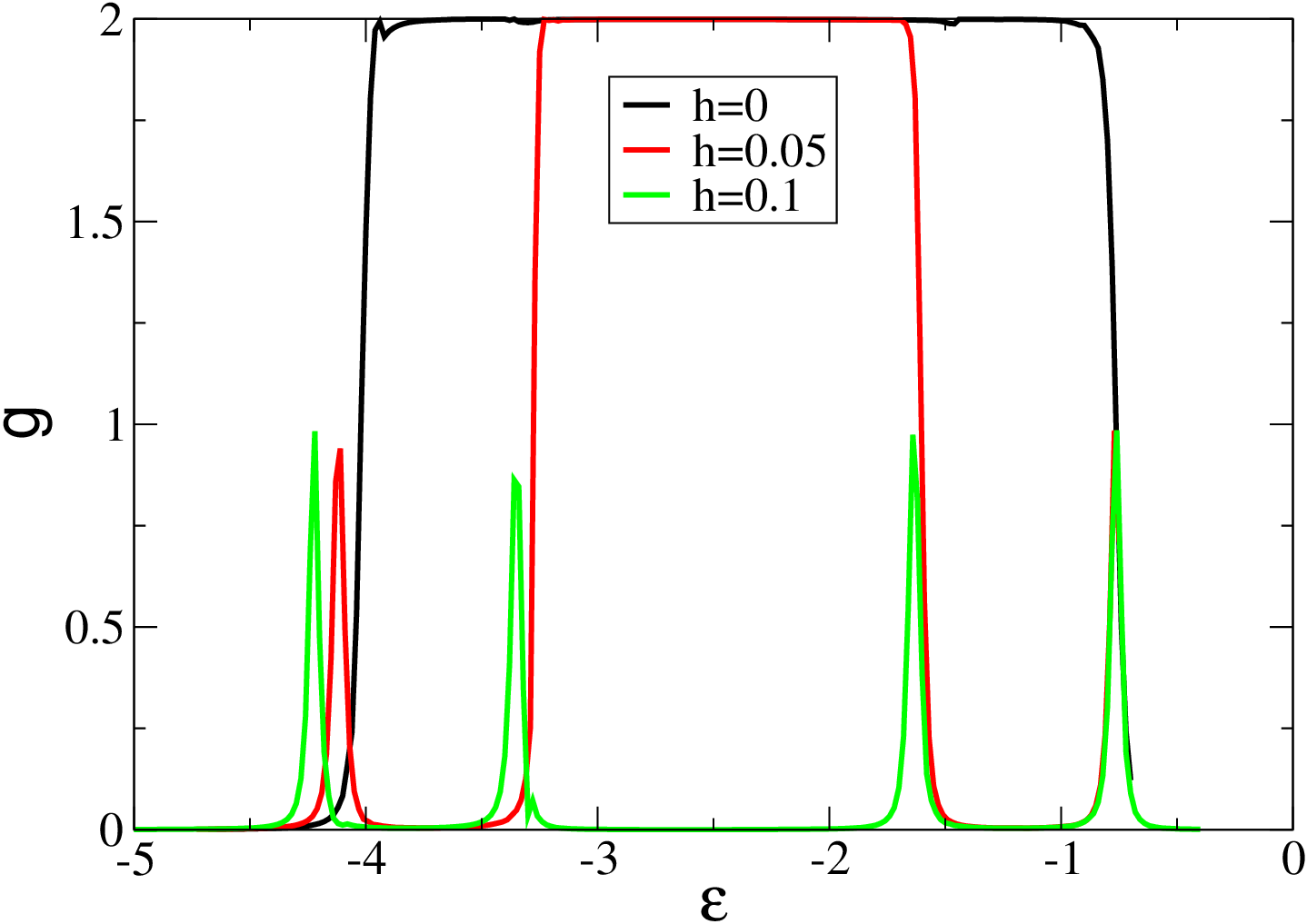} \vskip -.3truecm
\caption{(upper panel) Conductance at the middle of the
valley for $U=0.8$ and $V=0.05$: as function of
$t$ while $U^\prime=0$ (left) and as function of $U^\prime$ with $t=0.2$ (right).
The symbols corresponding to the numerical results fit quite well the line 
representing Eq. (\ref{con_mol}).
(lower panel) Conductance as function of $\epsilon$ for
$V=0.1$, $U=2$, $U^\prime=0.4$ and $t=0.23$ and no external magnetic field 
compared to the conductance in the presence of a magnetic field $h=0.05$.
With no magnetic field $g \sim 2$ for all molecule populations ranging between
$N=1$ and $N=3$. When a weak magnetic field is applied $g \sim 2$ only for $N=2$,
while for a stronger field $4$ Coulomb blockade peaks are observed.
} \label{fig2}
\end{figure}

The mechanism behind the conductance plateau of the $N=2$ state of the molecule
has to do with the influence of the interaction on the ground state of the $N=2$
sector. In the absence of interaction, the $+$ state ($|\uparrow \downarrow,0\rangle$)
is full and thus can not conduct, while the $-$ state ($|0,\uparrow \downarrow\rangle$)
is empty and thus also doesn't conduct. Due to interactions the molecule finds it
favorable to occupy a superposition of both states and therefore neither of them
are empty, and both, in principle may carry current. Since both states have an
$-\pi$ transmission phase difference, there will be interference between the two 
paths, which is taken into account in the  Friedel sum conductance
$g=\sin^2(\pi(n_{+\uparrow}-n_{-\uparrow}))+\sin^2(\pi(n_{+\downarrow}-n_{-\downarrow}))$.
Thus when $n_{+}=n_{-}$ destructive interference will occur and the conductance will be 
zero, while when $|n_{+}-n_{-}|=1/2$ maximum conductance per spin of $e^2/h$ is obtained.
Plugging in the values of  $n_{+}$ and $n_{-}$ as function of $t$, $U$ and $U^\prime$
previously calculated, we obtain the following expression for the valley plateau
dimensionless conductance:
\begin{eqnarray}
g=2 \sin^2 \left( \pi (1-[\sqrt{1+[4t/(U-U^\prime)]^2}-4t/(U-U^\prime)]^2)
\over 1+[\sqrt{1+[4t/(U-U^\prime)]^2}-4t/(U-U^\prime)]^2 \right). 
\label{con_mol}
\end{eqnarray}
This prediction was tested by comparing it to the numerical results for the conduction
in the middle of the valley, depicted in the upper part of Fig. \ref{fig2}. A reasonable
agreement can be seen. According to Eq. (\ref{con_mol}) the conductance in the valley
will be equal to its maximal value of $2$ once the argument of the $\sin$ is equal
to $\pi/2$, i.e., $n_{+}-n_{-}=1/2$. Thus, for
\begin{eqnarray}
t={{U-U^\prime}\over{4 \sqrt{3}}},
\label{con_max}
\end{eqnarray}
the valley conductance should be equal to $2$ (if the molecule is asymmetricaly
coupled to the leads, i.e., $V_L \ne V_R$ the valley height, as well as the
Kondo peak height, will be $8 (V_L V_R)^2/(V_L^2 + V_R^2)^2$). 
As can be seen in the lower panel in
Fig. \ref{fig2}, for a double dot in which $t$ is tuned to the value expressed
in Eq. (\ref{con_max}), the conduction for all molecule populations ranging between
$N=1$ and $N=3$ is equal to $2$. The conductance value of $g \sim 2$ stems from
two different processes. In the $N=1$ and $N=3$ regime it is the usual Kondo conductance
while in the $N=2$ it stems from the constructive interference between the
conductance through the $+$ and $-$ states. This can be clearly seen
when an external magnetic field is applied. The Kondo conductance is then
suppressed, but the $N=2$ state conductance isn't. 

Temperature is expected to have a very different influence
on the conductance for the different
sectors of the gate voltage. As long as the molecule is in the Kondo regime
(i.e., $N=1,3$), the relevant energy scale is the Kondo temperature, $T_K$,
which depends exponentially on the tunneling coupling of the molecule to the
lead ($V$). On the other hand, in the ``valley'' ($N=2$) as long as 
inelastic effects are ignored (in 1D systems both electron-electron 
and electron-phonon scattering are rather weak) the conductance
will change significantly only when $k_B T$ is of order of the singlet-triplet separation
$k_B T \sim (U^\prime-U)/2+\sqrt{(U-U^\prime)^2+(4t)^2}/2$.

In conclusion, the conductance through a strongly bound
identical double dot molecule was studied. The
gate voltage applied on the molecule changes its filling  
and therefore changes its conductance. Since the molecule
is relatively weakly coupled to the leads, the number of
electrons on the molecule is a good quantum number. For low temperatures ($T<T_K$) odd
fillings ($N=1,3$) lead to the usual Kondo conductance of $2e^2/h$. 
For the even fillings of $N=0,4$ the conductance is equal to zero.
The conductance for $N=2$ is sensitive to the parameters of the molecule, and
can assume values between $0-2e^2/h$. This behavior
stems from the partial filling of both the symmetric and anti-symmetric
orbitals due to interactions.
Thus current fllows through both orbitals and the conductance is the result
of the interference between them. Thus, as function
of the gate voltage a wide resonance tunneling plateau in the conductance appears,
which is quite unsensative to temperature and applied magnetic field.

We thank the Max Plank institute in Dresden where part of the work was
performed for its hospitality. RB thanks Y. Gefen, F. von Oppen, A. Schiller
and H. A. Weidenm\"uler for very usefull discussions and the Israel Academy of Science (Grant 877/04) for support.

\end{document}